\documentclass[12pt]{article}
\usepackage{latexsym,colordvi,dsfont}
\usepackage{graphicx}
\newcommand{\dslash}{\partial\hspace{-.2cm}/\hspace{.05cm}}
\newcommand{\lesim}{${\lower 2pt\hbox{$\scriptstyle
<$}\atop\raise 4pt\hbox{$\scriptstyle\sim$}}$} 
\begin{document}
\begin{center}
\begin{flushright}
     SWAT/02/359\\
December 2002
\end{flushright}
\vskip 3.5cm
{\Large \bf
Mass Generation without Phase Coherence at Nonzero Temperature 
}
\vskip 2.8 cm
Costas G. Strouthos and David N. Walters
\vskip 0.3 cm
{\em Department of Physics, University of Wales Swansea,\\
Singleton Park, Swansea SA2 8PP, U.K.}
\vskip 2.0 cm
\end{center}
\noindent{\small
%{\bf Abstract:} 
We present results from numerical simulations of the 2+1$d$
$SU(2)\otimes SU(2)$ Nambu--Jona-Lasinio model with $N_f=4$ fermion
flavors at zero and nonzero
temperature $T$. 
At zero temperature, critical exponents extracted from the scaling of the
order parameter and fermion mass are
found to be consistent with next-to-leading order predictions of the 
$1/N_f$ expansion.
At nonzero temperature we observe fermion mass generation despite the
lack of chiral symmetry 
breaking, which is forbidden by the Colemann-Mermin-Wagner theorem for
all $T>0$. We study the effects of lattice discretisation and finite volume 
on the dynamically generated fermion mass. By studying the lattice
dispersion relation we also show that in the hot phase there is no
significant temperature induced modification to the speed of light.
Studies of the equation of state are made by measuring
the pressure as a function of temperature and comparison is made with
large-$N_f$ predictions.
}

\newpage

\section{Introduction}

The nature of QCD and related field theories at finite temperature
remains a highly active area of both analytic and numerical
research. By studying the high temperature transition in strongly interacting theories we gain insight not only into the hot chirally symmetric
phase, but also the cold phase, in which spontaneous
chiral symmetry breaking leads to a dynamically generated
fermion mass.

In $2+1$ dimensional field theories the nature of the transition can be
particularly interesting.
It has been shown \cite{EB2,GN3U1} that in the three-dimensional Gross-Neveu model 
with a $U(1)$ chiral symmetry at nonzero temperature, there is a regime 
at temperatures above the Berezinzkii-Kosterlitz-Thouless transition in which the fermions
acquire nonzero mass despite the absence of phase coherence.  
An analogous phenomenon is found in the physics of strong
coupling or low carrier density superconductors. In these materials
Cooper-pair formation occurs below a
temperature $T=T^*$ whilst the $U(1)$ symmetry remains
manifest. The ``local" gap modulus which is neutral under $U(1)$
rotations is nonzero, whilst the phase of the gap fluctuates
violently \cite{EB1}; this is known as the pseudogap phase. It is only 
below a temperature $T=T_c\ll T^*$ that phase fluctuations cease
and the material enters a true superconducting phase. 
This separation of the temperatures of pair formation and pair condensation
has been known of for many years. 
It remains an open question, however, whether the dynamic
generation  of  fermion masses without the breaking of chiral
symmetry could be phenomenologically relevant in particle
physics.

In this paper we present a thermodynamic study of the $2+1d$
$SU(2)\otimes SU(2)$ Nambu$-$Jona-Lasinio (NJL) model on the
lattice, with the aim of showing that these phenomena can occur in
theories with more phenomenologically motivated symmetries.
 
In the continuum, the model is described by the Euclidean Lagrangian density 
\begin{equation}
{\cal L}  = \overline{\psi}_i\dslash\psi_i - \frac{g^2}{2N_f} 
 \left[\left(\overline{\psi}_i\psi_i\right)^2-\left(\overline{\psi}_i 
\gamma_5 \vec\tau\psi_i\right)^2\right], 
\label{Lcont}
\end{equation}
where $\vec{\tau}\equiv(\tau_1,\tau_2,\tau_3)$ are the Pauli spin
matrices which run over the internal $SU(2)$ isospin symmetry and are
normalised such that ${\rm t r}(\tau_\alpha\tau_\beta)=2\delta_{\alpha\beta}$.
The index
$i$ runs over $N_f$ fermion flavors and $g^2$ is the coupling
constant of the four-fermi interaction. 

The model is chirally
symmetric under $SU(2)_L\otimes SU(2)_R$;
$\psi\to (P_LU+P_RV)\psi$,
where $U$ and $V$ are independent global $SU(2)$ rotations and the
operators 
$P_{L,R}\equiv\frac{1}{2}(1\pm\gamma_5)$
project onto left
and right handed spinors respectively. It is also
invariant under $U(1)_V$
corresponding to a conserved baryon number.

The theory becomes considerably easier to
treat, both analytically and numerically, if we 
introduce scalar and pseudo-scalar
fields denoted by $\sigma$ and $\vec\pi$ respectively. 
The bosonised Lagrangian is
\begin{eqnarray}
{\cal L} & = & \overline{\psi}_i\left(\dslash+\sigma+i\gamma_5
\vec\pi \cdot \vec\tau\right)\psi_i
+\frac{N_f}{2g^2}\left(\sigma^2+\vec\pi \cdot \vec\pi\right) \nonumber \\
& = &  \overline{\psi}_i\left(\dslash+\sigma+i\gamma_5
\vec\pi \cdot \vec\tau \right)\psi_i
+\frac{N_f}{4g^2}{\rm tr}\Phi^\dagger\Phi,
\end{eqnarray}
where the combination $\Phi\equiv\sigma+i\vec\pi \cdot \vec\tau$ is
proportional to an element in the chiral group such that the model is
invariant under the rotation
$\Phi\to V\Phi U^{-1}$.

Apart from any obvious
numerical advantages this relatively simple model has various
interesting properties:
\begin{itemize}
\item[(i)] {the spectrum of excitations contains distinguishable
baryons and mesons, i.e. the elementary fermions $q$ and the composite
$q\overline{q}$ states;}
\item[(ii)] {at zero temperature and sufficiently strong coupling 
$g^2 > g^2_c$, chiral symmetry is spontaneously broken
leading to a dynamically generated fermion mass
equal to $M(0) = \langle \sigma \rangle$ in the large-$N_f$ limit;
the pion fields are the associated Goldstone bosons;}
\item[(iii)] {for dimensionality $2<d<4$ there is an interacting
continuum limit at a critical coupling, which for $d=3$ has a numerical value
$g_c^2/a\approx 1.0$ to leading order in $1/{N_f}$ with a lattice
regularisation employed;}
\item[(iv)] {the global symmetries are the same as those of 2 flavor
QCD, for which reason its four-dimensional version has a long history
as an effective theory of QCD at intermediate energies.}
\end{itemize}

Having highlighted the model's zero temperature properties, let us now
discuss what we know of its thermodynamics.
To leading order in $1/N_f$ the effective potential has the same form as 
the discrete symmetry case with the replacement $\sigma^2 \rightarrow \sigma^2 + \vec{\pi}^2$.
This implies that
chiral symmetry remains broken up to a critical temperature
$T_c=\frac{M}{2 \ln2}$ ($M$ is the zero temperature fermion mass)
and that symmetry restoration at this temperature is associated with
the fermions becoming massless.  
This conclusion is expected to be valid only when $N_f$ is strictly infinite,
i.e. when the fluctuations of the bosonic fields are neglected, since otherwise
it runs foul of Coleman-Mermin-Wagner (CMW) theorem \cite{CMW} which states 
that in $2+1$ dimensional systems a continuous symmetry must be manifest 
for all $T>0$. 
This behaviour may be understood by noting that, in the language of
spin models,  a domain wall separating
regions of oppositely oriented magnetisation has an energy which 
does not increase with the size of the system. If a wall has length 
$l$ and thickness $t$, the energy is $\sim l/t$, since the magnetisation
is allowed to interpolate continuously \cite{CMW}. This remains finite
even as $l, t \rightarrow \infty$. The physics is comparable, therefore, 
to that of the Ising chain, where a kink also costs only finite energy. 
At $T=0^+$ the magnetisation changes discontinuously, but there is no
latent heat as the domains can grow to be very large. A simple estimate
of the entropy \cite{CMW} reveals that the domain size grows
exponentially with the inverse temperature. Studies of the critical
properties of the $2+1d$ NJL models 
with  $Z_2$ \cite{GN3Z2} and $U(1)$ \cite{EB1,EB2,GN3U1}
chiral symmetries at nonzero temperature  
have shown that the thermally induced phase transition of the 
$Z_2$-symmetric model belongs to the universality class of the
two-dimensional Ising model and that the $U(1)$-symmetric model has
the same phase structure as  
the two-dimensional $X Y$ model. 
These results are in accordance with the dimensional reduction scenario, 
which predicts that the long range behaviour at the chiral phase transition 
is that of the $(d-1)$ spin model with the same symmetry. This is
because the IR 
region of the system is dominated by the zero Matsubara mode of the
bosonic field. 

In the $SU(2) \otimes SU(2)$-symmetric model, one
expects to observe fermion mass generation in the absence of symmetry
breaking at finite temperature, as in the case of the $U(1)$ model.
The local amplitude 
$\rho = \sqrt{\sigma^2+\vec\pi\cdot\vec\pi}$ is neutral under 
$O(4)$ rotations and may therefore be nonzero without breaking the symmetry. 
Hence, it is possible to have a dynamically generated fermion mass
$M(T) \sim \rho$ whose value may be comparable to the naive prediction 
of the large-$N_f$ approach. This was 
demonstrated in $d=(1+1)$
in \cite{Witten}, in which results from an exactly soluble fermionic 
model were generalised to the $U(1)$-symmetric Gross-Neveu model. As
was argued in \cite{GN3U1}, 
at $T\neq0$ the fermionic spectral function is modified to a branch cut, 
implying that the propagating fermion constantly emits and absorbs 
massless scalars and hence has an indefinite chirality. The physical
fermion may then be a superposition of positive and negative chirality
and therefore be neutral under chiral transformations.
It is in this
way that one anticipates mass generation despite manifest symmetry.

In this study we have carried out lattice simulations with
$N_f=4$. After briefly discussing the lattice formulation of the
model,  we present results of simulations at $T=0$ from which we
have extracted the bulk critical exponents $\beta_{\rm mag}$ and
$\nu$. For $T\neq0$ we have attempted to study the issue of mass
generation by measuring the so-called
screening or ``spatial'' mass $M_s(T)$ and the pole or ``temporal''
mass $M_t(T)$. In order to better understand the relationship between the
two we carried out a study of the lattice dispersion
relation. 
Finally, simulation on lattices with various temporal extents helps to
illustrate the nature of finite size and lattice discretisation effects.
Measuring the  pressure and making comparison with the
Stefan-Boltzmann limit was found to elucidate the matter further.

\section{Lattice Formulation}
\label{Formulation}
In its bosonised form, the model may be formulated on the lattice using
the action
\begin{equation}
\label{action}
S=\sum_{i=1}^N\sum_{x} \left[\overline{\chi}_i M[\sigma,\vec\pi]\chi_i
        +\overline{\zeta}_iM^*[\sigma,\vec\pi]\zeta_i\right] + 
\frac{2N}{g^2}\sum_{\tilde{x}}\left(\sigma^2+\vec\pi.\vec\pi\right),
\end{equation}
where $\chi$, $\overline{\chi}$, $\zeta$ and $\overline\zeta$ each
represent $N$
independent Grassmann-valued staggered fermionic fields defined on
lattice sites $x$, and the auxiliary 
bosonic fields $\sigma$ and $\vec\pi$ are defined on dual lattice
sites $\tilde{x}$. 
The fermion kinetic operator is given by
\begin{eqnarray}
M_{x y}^{p q}[\sigma,\vec\pi] & = &\frac{1}{2}\delta^{p q}
\left[\left(\delta_{y x+\hat{0}}-\delta_{y x-\hat{0}}\right) +\sum_{\nu=1,2} 
\eta_\nu(x)\left(\delta_{y x+\hat{\nu}}-\delta_{y x-\hat{\nu}}\right)\right] 
\nonumber \\
 &+& \frac{1}{8}\delta_{x y}\sum_{\left<\tilde{x},x\right>}
\left(\sigma(\tilde{x})\delta^{p q}+
i\epsilon(x)\vec\pi(\tilde{x}).\vec\tau^{p q}\right),
\end{eqnarray}
where $\left<\tilde{x},x\right>$ represents the set of 8 dual
lattice sites neighbouring $x$,
and the symbols $\eta_\nu(x)$ and $\epsilon(x)$ are the phases
$(-1)^{x_0+\ldots+x_{\nu-1}}$ and $(-1)^{x_0+x_1+x_2}$
respectively. $p,q$ are the internal $ SU(2)$ isospin indices, which
we include explicitly.
Making the replacement $\gamma_5\to\epsilon(x)$, the symmetries
defined in the previous section are
still observed, i.e. the $SU(2)\otimes SU(2)$ chiral symmetry remains
exact \cite{HM}, which is not the case in lattice QCD. With this
replacement, the projection operators $P_{L,R}\to P_{e,o}$
now project onto even and odd sub-lattices respectively.

This choice of action 
corresponds to using a functional weight of $\det{M^\dagger M}$. 
This doubles the fermionic degrees of freedom, but does allow one to
perform simulations using a hybrid Monte Carlo algorithm, which has the advantage
of being exact. Further doubling due to the use of staggered
fermions in three dimensions leads to a total of $N_f=4N$ continuum
fermion flavors. 
Trajectory lengths were sampled from a Poisson distribution with a mean
of approximately 1.0. This ensures
the ergodicity of the ensemble and reduces the autocorrelation of
successive configurations.

Simulations were
carried out on lattices with $L_s^2\times L_t$ sites, separated by
lattice spacing $a\sim1/\Lambda$, where $\Lambda$ is the ultra-violet
cut-off. 
In Euclidean spacetime, temperature is identified with the inverse temporal
extent, i.e. $T=1/(L_t a)$.
The temperature of the system may be altered, therefore,  by
varying either $\beta\equiv1/g^2$, which amounts to varying 
the lattice spacing, or by varying $L_t$. The effective zero
temperature limit is 
reached if one chooses $L_t\simeq L_s\gg 1$ and cut-off dependent
effects may be explored by varying $\beta$. 
To reach the continuum limit one has to satisfy the following two 
conditions: $\Lambda \gg T$ and $\Lambda \gg m$, where $m$ is a  mass scale
inversely associated with the correlation length $\xi$ of the
fluctuations of the order
parameter. The condition $\Lambda \gg T$ requires a lattice
with sufficiently large $L_t \gg 1$. 
This condition not being met results in the sum over the Matsubara frequencies
being truncated at too small a value. 
The second condition, 
$\Lambda \gg m$, implies that the coupling $\beta$ should be tuned toward
its critical value at which the correlation length $\xi \gg 1$.
Furthermore, the condition $L_s \gg \xi$ should be satisfied in order
to minimise finite size effects and obtain  results which approximate
the thermodynamic limit. 

\section{Results}
\subsection{Zero Temperature}
In order to study the chiral symmetry of the model first at $T=0$ and
then at $T>0$ we work in the
chiral limit, i.e. we choose not to introduce a bare fermion mass $m_0$ into
eq.~(\ref{action}). 
 Without the benefit of this interaction, 
the direction of symmetry breaking changes over the course of the simulation
such that $\Sigma\equiv\frac{1}{V}\sum_{\tilde{x}}\sigma(\tilde{x})$ and 
$\vec\Pi\equiv\frac{1}{V}\sum_{\tilde{x}}\vec\pi(\tilde{x})$ (where $V$ is the lattice
volume) average to zero over 
the ensemble. It is in this way that the absence of spontaneous
chiral symmetry breaking on the finite lattice is enforced. 
Instead, we define an effective order parameter   
\begin{equation}
\Phi\equiv\left<\sqrt{\Sigma^2+\vec\Pi \cdot \vec\Pi}\right>,
\end{equation} 
which is a
projection onto the direction of symmetry breaking for each
configuration. Although this definition allows one to distinguish
between the chirally broken and unbroken phases, it should be noted
that the two definitions of the order parameter, $\Phi$ and
$\left.\left<\sigma\right>\right|_{m_0\to0}$, differ by a factor of
$1/\sqrt{V}$ as it
is impossible to 
disentangle the chiral order parameter field in $\Phi$ from the Goldstone
modes. 
We also measure the fermion correlation function for temporal
separation $t$; 
\begin{equation}
\begin{array}{c c c l}
C(t)=A_t\left[e^{-M t}+e^{-M(L_t-t)}\right]&&{\rm if}&t={\rm odd},\\
C(t)=0&&{\rm if }&t = {\rm even},
\label{C(t)}
\end{array}
\end{equation}
which allows one to extract the fermion
pole mass $M$. 
The fact that $C(t={\rm even})$ averages to zero over the
ensemble reflects the fact that chiral symmetry is not truly broken on
our finite size system, as manifest chiral symmetry implies that
the $G_{o o}$ and $G_{e e}$ components of the staggered fermion
propagator should vanish.  

\begin{figure}[ht]
\centering
\includegraphics[width=11cm]{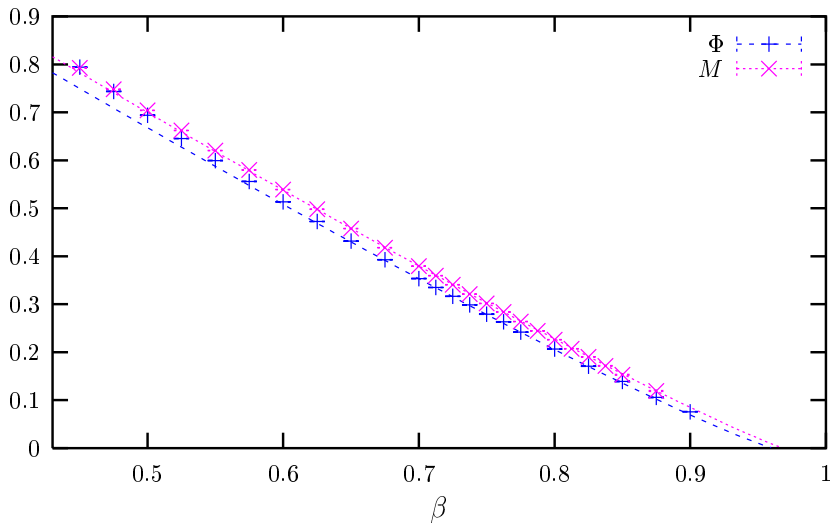}
\caption{Effective order parameter and fermion mass as functions of
$\beta$ on $36^3$ and $48^3$ lattices.}
\label{T=0}
\end{figure}

We simulated the model on a $36^3$ lattice with 
repeated simulations on a $48^3$ lattice near the 
critical coupling $\beta_c^{\mathrm{bulk}}$
in order to detect and control finite size effects.
Approximately 1000 equilibrated trajectories
were generated for each point. Fitting these results to the scaling
forms $\Phi=a_1(\beta^{\rm 
bulk}_c-\beta)^{\beta_{\rm mag}}$ and $M=a_2(\beta^{\rm bulk}_c-\beta)^\nu$
we were able to extract the critical exponents $\beta_{\rm mag}$ and
$\nu$. As $\beta\to 0$, 
the space-time lattice becomes large and coarse
and the ${\cal O}(a)$ lattice discretisation effects become
significant. As $\beta\to \beta^{\rm bulk}_c$ the decreasing lattice
spacing causes the lattice to become less coarse, but finite volume
effects become 
dominant. True scaling behaviour is limited, therefore, to a
particular window, which we choose as $0.650\leq\beta\leq0.825$ by
demanding that results of the fits be stable with respect to the
adding and subtracting of 
individual points. The measured values of the exponents
were $\beta_{\rm mag}=1.12(2)$ and $\nu=1.10(2)$ with $\beta_c^{\rm
bulk}=0.961(5)$ and $0.969(3)$ respectively. 
Data for $\Phi$ and $M$ versus $\beta$ are plotted 
with the fitted scaling functions in  Fig.~\ref{T=0}.

Our results for $\nu$ and $\beta_{\rm mag}$ are consistent with the
$1/N_f$ expansion results calculated in \cite{AnnPhys};
\begin{eqnarray}
\beta_{\rm mag}&=&1+\frac{3}{2}\left(\frac{4}{\pi^2}\right)\frac{1}{N_f}
+{\cal O}\frac{1}{N_f^2} \sim 1.15
\label{beta_mag} \\
\nu&=&1+\frac{4}{3}\left(\frac{4}{\pi^2}\right)\frac{1}{N_f}
+{\cal O}\frac{1}{N_f^2} \sim 1.14,
\label{nu}
\end{eqnarray}
where in our simulations $N_f=4$. The fact that we observed
next-to-leading order corrections, which were not observed in studies
of the $Z_2$ \cite{AnnPhys} and $U(1)$ \cite{GN3U1} models, is  because the
${\cal O}(1/N_f)$ corrections in eq.~(\ref{beta_mag}) and eq.~(\ref{nu}) are
slightly larger than in the models of lower symmetry. The fact that
there are more light mesons that  
mediate the interaction in the
$SU(2) \otimes SU(2)$-symmetric model
than in the $Z_2$ and $U(1)$ models may be the source 
for the
larger deviation from the mean field values, which are $\beta_{\rm
mag}=\nu=1$ for all three symmetries.  
More accurate values of the exponent could be extracted by performing detailed
finite size scaling analyses in the vicinity of the critical point.

In addition to the extraction of critical exponents, the 
studies presented here provide a length scale, namely the fermion mass $M$,
with which one can present nonzero temperature results in a cut-off
independent fashion. This is performed in the next section.

\subsection{Nonzero Temperature}

\label{Tneq0}

In order to study the model at nonzero temperature we performed simulations on
lattices with a constant temporal extent $L_t=4$ and varying spatial
volumes $V_s=30^2, 50^2, 100^2, 150^2$ and $220^2$. The temperature was
varied  continuously by varying $\beta <
\beta_c^{\rm bulk}\sim0.965$, at which point $a\to0$ and $T\to\infty$.

\begin{figure}[ht]
\centering
\includegraphics[width=11cm]{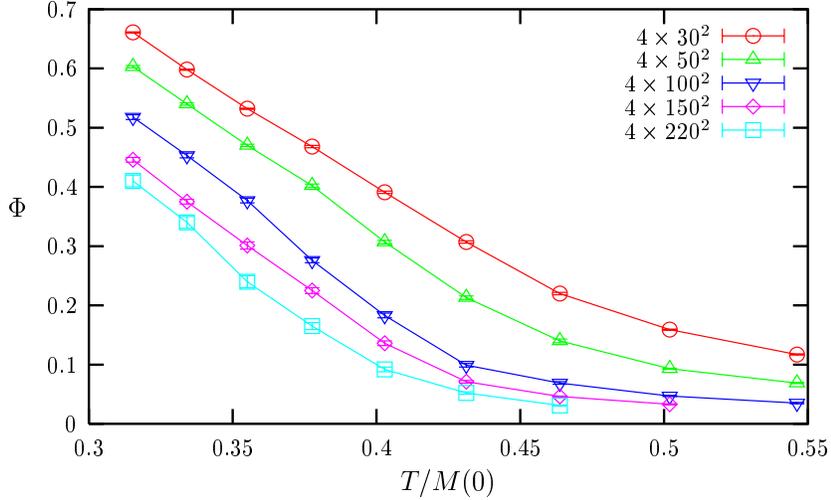}
\caption{Effective order parameter vs. normalised
temperature for various spatial volumes.}
\label{order_param_Lt_4}
\end{figure}

The effective order parameter is plotted against temperature in
Fig.~\ref{order_param_Lt_4}. The temperature is normalised by a factor of
the zero temperature pole mass $M(0)$, in order to make results cut-off independent as previously discussed. 
At high temperature, $\Phi$ is consistent with zero
whilst at low temperature there are significant finite size
effects. 
These effects are a signal of the correlation 
 length diverging as one approaches the transition at $T=0^+$ and the
condition $L_s\gg\xi$ not being met. Once $\xi\sim L_s$ one is
unable to observe a manifest symmetry.
Our results are consistent with the expectation of chiral symmetry restoration
in the infinite volume limit for $T\neq0$.

\begin{figure}[ht]
\centering
\includegraphics[width=11cm]{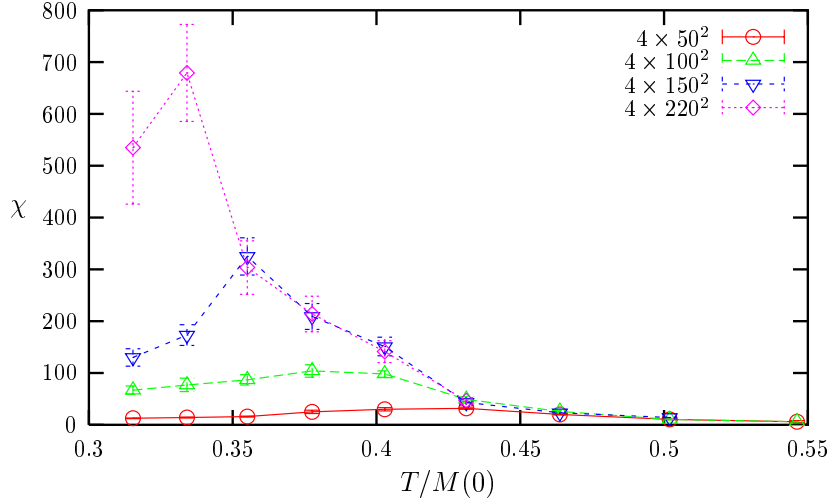}
\caption{Susceptibility vs. normalised
temperature for various spatial volumes.}
\label{suscept_LT_4}
\end{figure}

This picture is supported by the results shown in
Fig.~\ref{suscept_LT_4}, in which the susceptibility of the order
parameter
\begin{equation}
\chi=V\left(\left<\left|\Phi\right|^2\right>-\left<\left|\Phi\right|\right>^2\right),
\end{equation} 
is plotted as a function of temperature. As
spatial volume is increased, the pseudo-critical temperature
$T_c(L_s)$, i.e. the position of the peak in susceptibility, 
decreases, supporting a transition at $T=0^+$ in the infinite volume
limit. Finite size effects are again
very large in the  low $T$ regime,  which is again
consistent with the expectation that  the correlation length diverges
as $T \rightarrow 0$. It is elucidating to compare our results for
$\Phi$ and $\chi$ with those from simulations of the
$U(1)$-symmetric model \cite{GN3U1}, where it was shown that a
Berezinskii-Kosterlitz-Thouless transition separates two chirally
symmetric phases, one critical and one with finite correlation length.
The main observations are:
(a) in the $U(1)$ model the finite size effects in the low $T$ phase
are larger than in the $SU(2) \otimes SU(2)$ model because
the correlation length is infinite for any $T \leq T_{B K T}$; (b) the
positions of the  susceptibility peaks in the $U(1)$ model converge to
a nonzero $T=T_{B K T}$. 

In order to address the issue of mass generation with $L_t=4$, we cannot 
directly extract the fermion mass from temporal correlators via
eq. (\ref{C(t)}) as there are no longer enough temporal
separations over which to carry out a fit.
 Instead, by studying 
the decay of spatial correlators, we extract the so-called screening 
or ``spatial'' mass $M_s(T)$ via
\begin{equation}
\begin{array}
{c c c l}
C(s)=A_s\left[e^{-M_s s}+e^{-M_s(L_s-s)}\right]&&{\rm if}&s={\rm odd},\\
C(s)=0&&{\rm if }&s = {\rm even}.
\label{C(s)}
\end{array}
\end{equation}
For free, massless fermions we expect this quantity to be the lowest
fermion Matsubara 
mode, given in the continuum by $\omega_0^{\rm cont}=\pi T$, and on the
lattice by $\omega_0^{\rm lat}=\sinh^{-1}\left[\sin(\pi/L_t)\right]$; $\omega_0^{\rm
lat}\approx 0.658$ for $L_t=4$. 
In the low temperature phase of the large-$N_f$ limit the ``temporal'' or 
pole fermion mass $M_t(T)$ is related to the screening or ``spatial''
mass $M_s(T)$ via $M_s^2=\omega_o^2+M_t^2$, as the effects of the
interaction are absorbed  by the dynamically generated mass. If
$1/N_f$ effects are taken into account, however, the relation between $M_t$
and $M_s$ can be less trivial.  In particular, at nonzero temperature the
fermionic dispersion relation is expected to reflect the breaking of
Lorentz invariance via a temperature dependent vacuum polarisation
term $\Pi(\vec{k},T)$: 
\begin{equation}
E^2(\vec{k},T)=M^2(0) +\vec{k}^2 +\Pi(\vec{k},T).
\label{disprel1}
\end{equation}
Temperature effects can be absorbed into the thermal mass $M(T)$ and a coefficient $A(T)$
which can differ from one,
implying a temperature dependent modification to the speed of light.
The latter effect is expected to be significant for $\left|\vec{k}\right| \ll T$,
allowing one to write
$\Pi(\vec{k},T)=\alpha^2(T)\vec{k}^2+{\cal O}(\vec{k}^4)$.
The dispersion relation becomes
\begin{equation}
E^2(\vec{k},T)=M^2(T) + A^2(T)\vec{k}^2,
\label{disprel2}
\end{equation}
where $A^2 = \alpha^2 + 1$. It is simple to show that the
relationship  
between the screening and the pole mass is given by
$A^2 M_s^2 = M_t^2 + \omega_0^2$, implying that a knowledge of $M_s$
alone is not sufficient to reach a conclusion about the magnitude of 
the dynamical fermion mass $M_t$.
This issue is investigated later in this section.
 
\begin{figure}[ht]
\centering
\includegraphics[width=11cm]{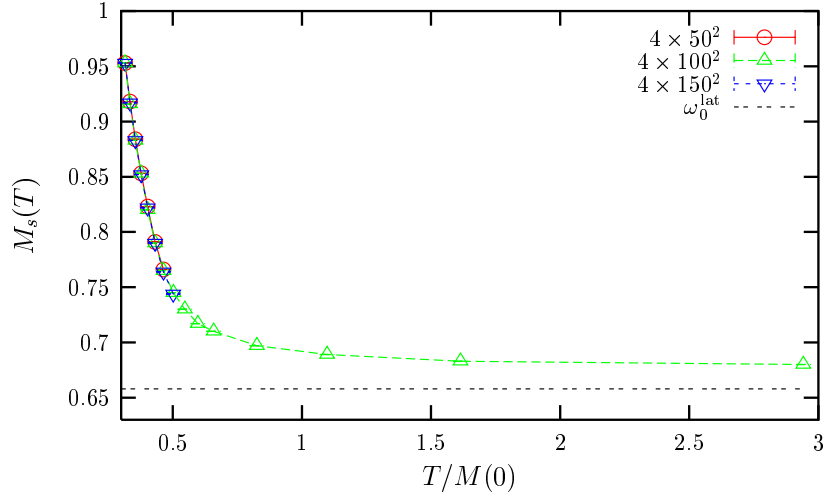}
\caption{Screening mass vs. normalised temperature. The horizontal line
represents the lowest Matsubara mode.}
\label{ms_Lt=4}
\end{figure}

Results from the measurement of the screening mass from the $L_t=4$
simulations on various spatial volumes are presented in
Fig.~\ref{ms_Lt=4}.
The first observation to be made is that at low temperature
$M_s(T)$ is significantly 
larger than $\omega_0^{\rm lat}$. It is an interesting aside to note
that, as was the case with the $U(1)$ model \cite{GN3U1}, the
magnitude of the screening mass in our simulations 
is comparable to the values extracted from 
simulations of the $Z_2$-symmetric model \cite{GN3Z2} where mass generation 
is ascribed to orthodox chiral symmetry breaking.
Fig.~\ref{ms_Lt=4} shows the mass to be size independent, 
even though, as we have seen from Fig.~\ref{order_param_Lt_4},
the order parameter $\Phi$ has
large finite size effects for all $T>0$.
It is also clear from Fig.~\ref{ms_Lt=4} that $M_s$
does not approach a value 
equal to $\omega_0^{\rm lat}$ in the limit that $T\to\infty$. 
\begin{figure}[ht]
\centering
\includegraphics[width=11cm,]{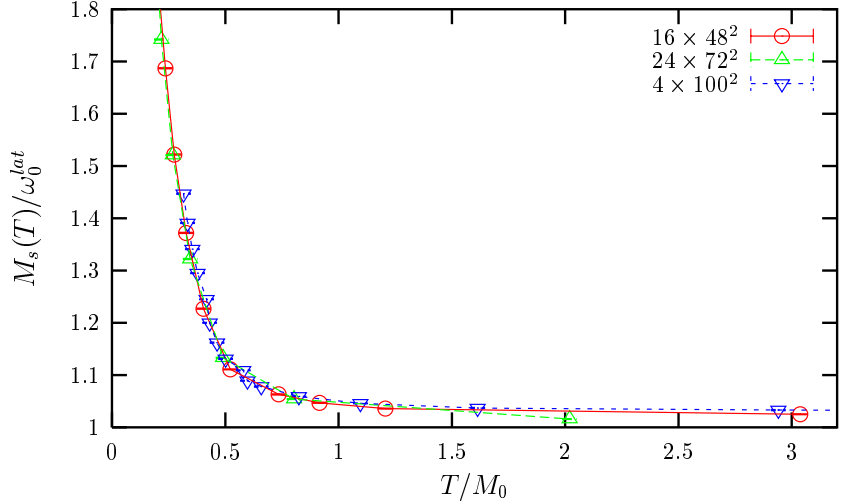}
\caption{Screening mass vs. normalised temperature. In order to
compare different $L_t$ we normalise by a factor of $\omega_0^{\rm lat}$. }
\label{ms_Ltneq4}
\end{figure}
We attempted to understand this by measuring the fermion mass on
lattices with different temporal extents $L_t=8$, $16$ and
$24$  and for a range of values of
the coupling $\beta$. 
 The screening mass measured on lattices with three different values of
$L_t$ are presented in 
Fig.~\ref{ms_Ltneq4}. The data are normalised by a factor of
$\omega_0^{\rm lat}$ to make it possible to plot the
results on the same axes. At low temperature there
seems to be no $L_t$ dependence whilst at high temperature we see that
the discrepancy between $M_s(T)$ and $\omega_0^{\rm lat}$ is reduced as
$L_t$ is increased. 
The effect of finite $L_t$ is that only a
finite number of Matsubara modes are present on the lattice, and
when the fermion screening mass becomes comparable
with the lowest Matsubara mode $M_s(T)\sim\omega_0^{\rm lat}$,
discretisation artifacts become significant. 
It
is clear that this effect is only apparent when
comparison is made with large-$L_t$ data.

\begin{figure}[ht]
\centering
\includegraphics[width=11cm]{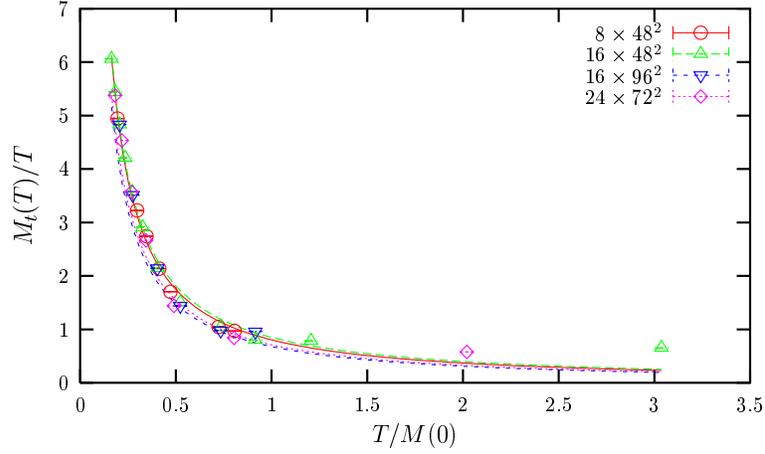}
\caption{$M_t(T)/T$ vs. $T/M(0)$.}
\label{pole.mass.2}
\end{figure}

In order to gain insight into the effects of finite $L_t$ and finite physical 
volume (which at nonzero
temperature is translated into finite asymmetry ratio $L_s/L_t$) we
studied the pole mass on different lattice sizes $8\times48^2$, $16\times48^2$,
$16\times96^2$ and $24 \times 72^2$.  
\begin{figure}[ht]
\centering
\includegraphics[width=11cm]{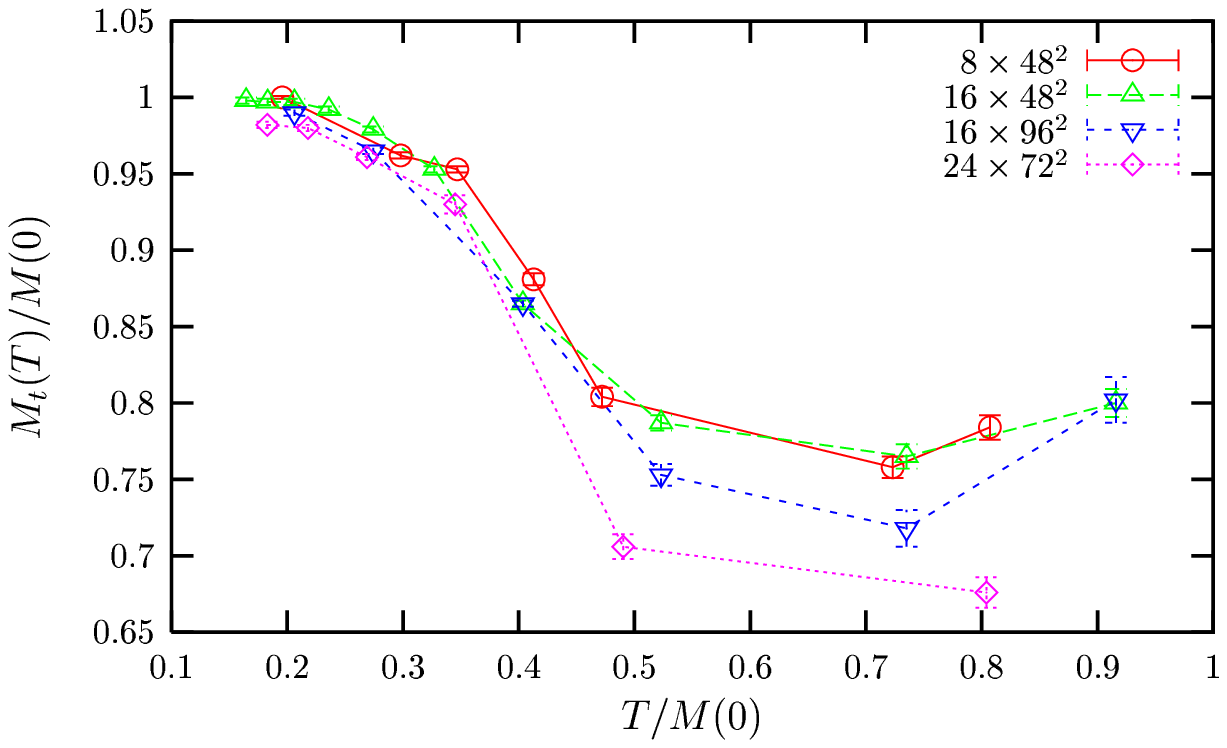}
\caption{$M_t(T)/M(0)$ vs. $T/M(0)$.}
\label{pole.mass.1}
\end{figure}
The results of the pole mass normalised with
respect to $M(0)$ are presented in Fig.~\ref{pole.mass.1}. 
At this point we remind the reader that in order to increase $T$, we decrease 
the lattice spacing $a$, which is achieved by tuning the coupling $\beta$
toward the bulk critical point $\beta_c^{\rm bulk}$.
Comparing curves with the same asymmetry ratios and different $L_t$,
such as those extracted from data on lattice sizes $8 \times 48^2$
with $16 \times 96^2$ and $16 \times 48^2$ with $24 \times 72^2$,  
we see that the finite $L_t$ effects become large as $T$ increases.
As in Fig.~\ref{ms_Ltneq4}, this is due to the fermion screening
mass becomes comparable with the lowest Matsubara mode.
Similarly, the effects of finite spatial extent become more pronounced
with temperature, since finite size effects become large
when the fermion correlation  length is comparable with  
the  spatial extent of the lattice $\xi\sim M_t^{-1}\sim L_s$.
We have plotted the same data with a different normalisation in
Fig.~\ref{pole.mass.2} and fitted to the form $M_t(T)/T=
\alpha_1\left(T/M(0)\right)^{\alpha_2}$. 
We chose to fit only to data with
$T/M(0)$\lesim$0.5$, as we know from Fig.~\ref{pole.mass.1} that 
in this region finite size and
discretisation errors are not severe. 
This choice is justified by noting that in Fig.~\ref{pole.mass.2} the
data at large $T$ deviate from the fitting line.
We found a weak 
temperature dependence, with $0.68<\alpha_1<0.85$ and 
$-1.25<\alpha_2<-1.12$ over the four fits. 
This is consistent with $M_t\to0$ as $T\to\infty$.
In the future, it would be interesting to compare this result with
next-to-leading order $1/N_f$ calculations.

As previously discussed, in order to better understand the temperature
dependence of eq.~(\ref{disprel2})
we have studied the fermion correlation function at 
nonzero momentum ${k}=2\pi n /L_s$  ($n=0,1,2,\ldots,L_s/4$) in the 
spatial direction and used the energy $E(k)$ extracted from 
eq.~(\ref{C(t)}) to map out the fermion dispersion relation. Due to the 
periodicity of the lattice in spatial dimensions and the 
doubling of fermion species, these dispersion relations are symmetric
around $k=\pi/2$.  We fitted data from the zero and nonzero 
temperature phases to the lattice free fermion dispersion relation \cite{Boyd}
\begin{equation}
\sinh^2E = A^2 \sin^2k +\sinh^2M_t. 
\label{disprel3}
\end{equation}
\begin{figure}[ht]
\centering
\includegraphics[width=11cm]{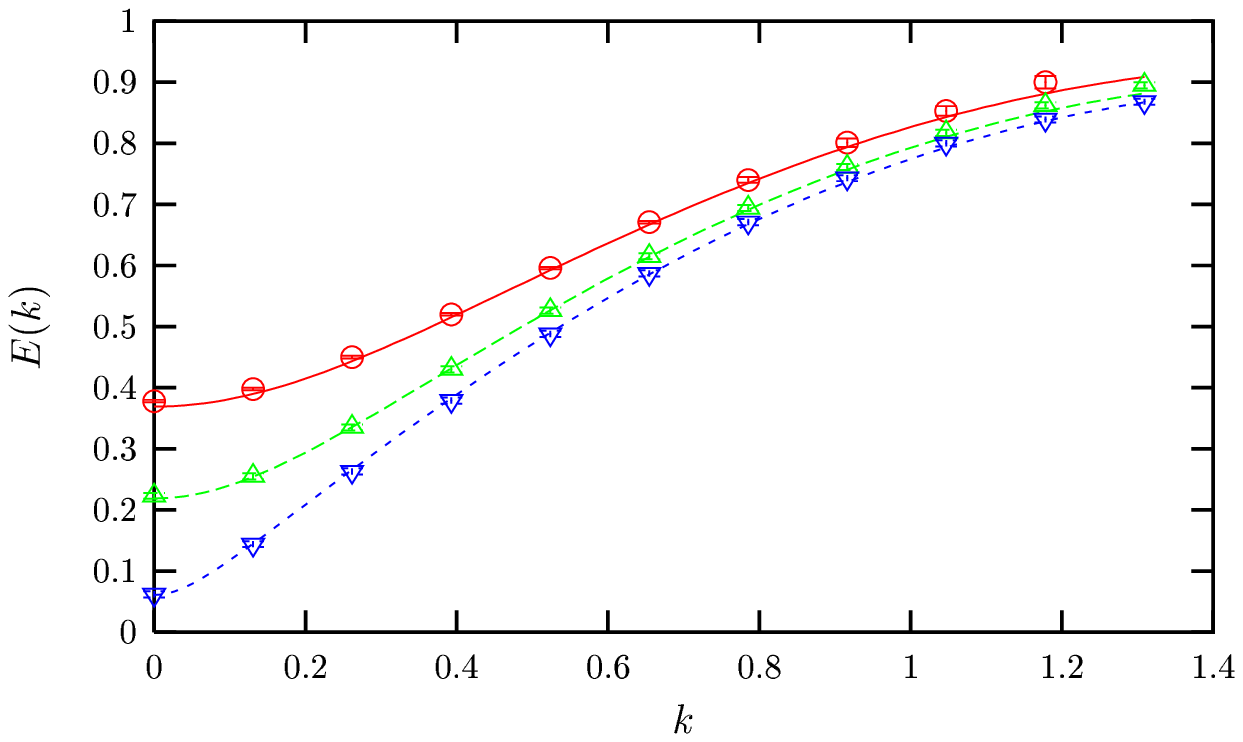}
\caption{Fermionic dispersion relation vs. momentum $k$.
The rescaled temperature $T/M(0)$ decreases from $0.165$ (top) 
to 0.735 (bottom).}
\label{Disp.Rel}
\end{figure}
Results from the hot phase on a $16\times48^2$ lattice are shown
with the fitted forms in Fig.~\ref{Disp.Rel}. 
The masses extracted from fits to $E(k)$ at both $T=0$ and $T>0$
are consistent with those
extracted from the zero momentum correlators and yield
values of $A\simeq1$ to within less than $3\%$. 
This implies that the principal physical effect of the hot medium is
to generate a nonzero thermal mass, rather than to renormalise the
speed of light.  
Furthermore, our results show that any extra factors not included in the 
free-field ansatz eq.~(\ref{disprel3}) are negligible.
Similar results were extracted by fitting the dispersion relation to
data from $16 \times 96^2$ and $24 \times 72^2$ lattices. 
The $A \simeq 1$ result provided additional evidence that the nonzero
screening mass extracted from our simulations with $L_t=4$ implies
that the fermion pole mass is nonzero. We see no modifications to the low
momentum part of the dispersion relation, expected to have
two branches at $T>0$ \cite{weldon}. These branches, corresponding to two types
of quasi-particle excitations (the fermion and the plasmino) are not
visible in our simulations.  

\begin{figure}[ht]
\centering
\includegraphics[width=11cm]{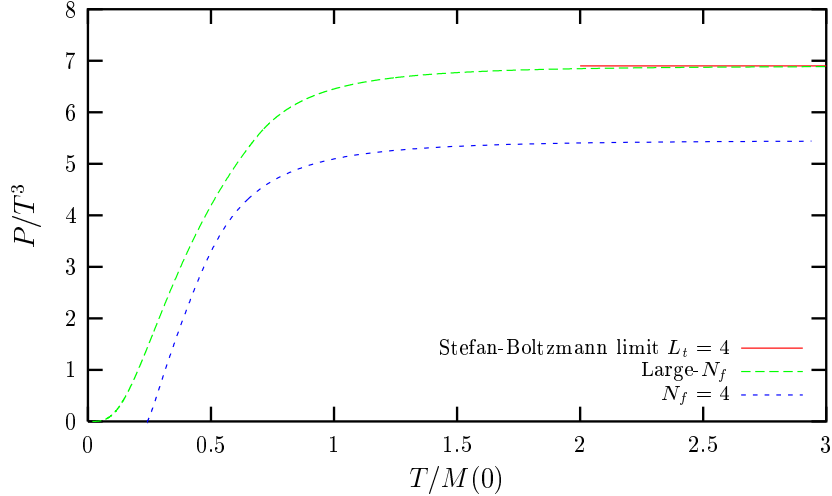}
\caption{The pressure calculated from the lattice gap equation
and from numerical simulations. In both cases the lattice size
is $4 \times 100^2$.}
\label{pressure_lt=4}
\end{figure}

Finally, to gain further insight into the thermodynamics 
of the model, we  studied the equation of state by 
measuring  the pressure as a function of temperature using the
integral method \cite{engels}. 
For homogeneous systems the pressure $P$ is given by 
\begin{equation}
P \equiv \left.\frac{\partial \ln {\cal Z}}{\partial V}\right|_T = \frac{\ln
{\cal Z}}{V}=-f, 
\end{equation}
where $f$ is the free energy density. 
On the lattice the logarithm of the partition function is calculated
from the expectation value of the bosonic part of the action 
$S_{b o s} \equiv 2N\sum_{\tilde{x}}[\sigma^2(\tilde{x}) + \vec{\pi}(\tilde{x}) \cdot \vec{\pi}(\tilde{x})]$.
Since the derivative with respect to the bare coupling $\beta$ is
\begin{equation}
-\frac{\partial Z}{\partial \beta} = \langle S \rangle,
\end{equation}
the physical free energy density can obtained from 
\begin{equation}
\label{free_energy}
f =\frac{1}{V} \int_{\beta_0}^{\beta} {\mathrm d}\beta^{\prime}\left[\langle S_{b o s} \rangle_0 - \langle S_{b o s} \rangle_T\right].
\end{equation}
The free energy density is normalised by subtracting the vacuum contribution calculated
at $T=0$. 
The pressure was calculated from data on the $4 \times 100^2$
lattice whilst the  
vacuum expectation value of $S_{b o s}$ was calculated from data
on $36^3$ and $48^3$ lattices. A lattice large-$N_f$ calculation of $P$,
including the calculation of the lattice Stefan-Boltzmann limit, is outlined
in Appendix A.
This result calculated on a $4\times100^2$ lattice is compared with
the result extracted from simulations with $N_f=4$ on the same lattice size
in Fig.~\ref{pressure_lt=4}. 
The statistical error in the
pressure extracted from simulation data is approximately $5\%$ of the
value of $P$ for all $T$. 
It is striking that the value of $P/T^3$ extracted from the
simulations doesn't reach the 
Stefan-Boltzmann limit as $T\to\infty$, an effect found to be
independent of the spatial volume. 
We believe that  this is simply another manifestation of
the effect that causes the screening mass not to approach
$\omega_0^{\rm lat}$ as $T\to\infty$. We interpret it as a
renormalisation of discretisation artifacts by $1/N_f$
corrections. Unfortunately, our data on lattices with $L_t > 4$
were not sufficient to study the equation of state because the fact
that as $L_t$ becomes large the two contributions to
eq.~(\ref{free_energy}) converge, in conjunction with the
normalisation factor of $T^{-3}=L_t^3$ causes statistical errors to
swamp any signal in the data.
 
Drawing the information in this section together, our results provide
clear evidence 
for the fermions acquiring nonzero mass in the absence of chiral
symmetry breaking. In order to be certain whether mass generation
switches off at some value of $T=T^*$ or whether it goes to zero
asymptotically at $T \to \infty$, we would need to simulate the model on
lattices with much larger temporal and spatial extents.

\section{Conclusions}

In this paper we have presented a thermodynamic lattice study of the $2+1$
dimensional $SU(2)\otimes SU(2)$-symmetric NJL model.
We find the model to have manifest chiral symmetry for all $T\neq 0$
in accordance with the CMW theorem. As temperature is reduced,
the susceptibility of the order parameter begins to diverge
and is consistent with there being a phase transition at $T=0^+$.
This should be contrasted with the
predictions of the large-$N_f$ limit, in which fluctuations of the
bosonic fields are neglected and chiral symmetry remains broken up
to a critical temperature $T_c=\frac{M(0)}{2\ln 2}$. In this
limit, the breaking of chiral symmetry is associated with the
generation of fermion mass in the standard way. We have
demonstrated non-perturbatively, that in agreement with
large-$N_f$ there is mass 
generation at $T\neq0$ in this model.

By studying the 
so-called ``screening'' mass $M_s(T)$,  we have shown that
at low but nonzero temperature $M_s\gg\omega_0$. With a temporal
extent of $L_t=4$, we observed that the spatial mass
fails to approach $\omega_0$ in the $T\to\infty$ limit. This effect is
also observed in the equation of state, where as $T\to\infty$,
the pressure fails to
approach the Stefan-Boltzmann limit.
This is due to the fact that when $L_t$ is small there
are only a small number of Matsubara modes available, an effect which
becomes noticeable at high temperatures, where $M_s\sim\omega_0$. 
Results from measurements of $M_s$ on 
lattices with $L_t>4$ have shown this to be a discretisation
effect which is reduced when $L_t\gg1$. 
By studying the fermion dispersion relation, we have also shown that
any temperature dependent modification to the relationship between
$M_s$ and $M_t$ is negligible, implying that we have unambiguous evidence
that at $T>0$ we observe mass generation despite the lack of chiral
symmetry breaking. 
It is interesting to note that the value of $M_s(T)$ is
of the same magnitude as it is in the model with $Z_2$ chiral
symmetry \cite{GN3U1}, in which mass generation is ascribed to conventional symmetry
breaking. 
To support our evidence of mass generation from the measurement of $M_s$
we have also measured the 
pole mass directly on lattices with $L_t\gg1$. At high temperatures we
observe large finite size 
effects as the correlation length of the low-mass fermion becomes
comparable to the spatial extent of the system. At low
temperatures, however, we observe that the mass moves smoothly toward
its zero temperature value.
As with the results from the $U(1)$ model, we take this as further
support for Witten's statement \cite{Witten} that when
interpreted correctly,  
the large-$N_f$ expansion is a reliable guide to the properties of the
model. Although the CMW theorem forbids the large-$N_f$ prediction of symmetry
breaking at $T>0$, the prediction of a dynamically generated fermion
mass is adhered to. 
In this case the propagating fermion 
constantly emits and absorbs massless scalars and hence has indefinite
chirality.  
As was argued in \cite{GN3U1},
the $T\neq0$ fermionic spectral function is modified to a branch cut.
In this model we observe mass generation without phase coherence,
i.e. we have a pseudogap phase for $T>0$. At this stage, however, we
cannot say whether there is a phase transition at some temperature $T^*$   
or the fermion mass goes asymptotically to zero as $T\to\infty$. In
order to study this regime (corresponding to $a\to0$) it is necessary
to perform new simulations 
on lattices with larger $L_t$ and $L_s$. 

To put these results into context, we note that the observation of
mass generation without chiral symmetry breaking in $2+1d$ adds to the
already rich and 
interesting phase diagram of the $SU(2)\otimes SU(2)$ NJL model
\cite{GN3PD}. At 
high baryon chemical potential and zero temperature a complementary
phenomenon is known to occur. Monte Carlo
simulations have shown there to be a ``thin film'' superfluid phase
with long range phase coherence, but no mass gap is generated about the
Fermi surface \cite{NJL3muneq0}. Although similar in nature to the
$2d$ $X Y$ spin model, 
a unique critical exponent seems to put the high-$\mu$ phase in a
new universality class.
In $3+1$ dimensions, the high-$T$ low-$\mu$ phase shows
conventional symmetry restoration via a $T\neq0$ transition and the
high-$\mu$ low-$T$ phase appears to be that of a traditional BCS
superfluid. A study of the region at high-$\mu$ and above
the critical temperature, however, shows it to behave as if in
a pseudogap phase \cite{Kitazawa}. Analogous to strong coupling
superconductors there
are two distinct critical temperatures: (a) $T^*$ below which there is
manifest baryon number symmetry but no mass gap $\Delta$, and (b) 
$T_c$ below which normal superfluidity occurs. This is particularly
interesting, as it provides evidence for mass gap generation without
phase coherence in a phenomenologically relevant number of dimensions.

\section*{Acknowledgements}
Costas G. Strouthos was supported by a Leverhulme Trust grant.
Discussions with Simon Hands are greatly appreciated.

\renewcommand{\theequation}{A-\arabic{equation}}
% redefine the command that creates the equation no.
\setcounter{equation}{0}  % reset counter 
\section*{Appendix A: Large-$N_f$ calculation of the Pressure}  
% use *-form to suppress numbering

In this appendix we outline a derivation for the
pressure in the large-$N_f$ limit on the lattice. 
In section \ref{Formulation} we define our model in terms of staggered fermion
fields which are defined on sites $x$ of the
space-time lattice. In $d$ dimensions, fermion doubling gives rise to
$2^d$ degrees of freedom per naive lattice flavor, which in the
staggered formalism are interpreted as $2^d/4$ 
``tastes''\footnote{We use the term ``tastes'' here to distinguish these
2 Kogut-Susskind flavors from our $N$ lattice and $N_f$ continuum
flavors.} of Dirac spinor. This is easier to understand 
if we redefine the fields on a
blocked lattice, with each site $y$ associated with $2^d$ sites of the
original lattice \cite{KlubergStern}. The kinetic operator then tends to 
the continuum
Dirac operator in the limit $a\to0$, with a pole in the propagator
only at zero momentum.
We apply this formalism, which is outlined for an odd number of dimensions in
\cite{Burden}, in the case that $d=3$.
Defining the $2\times2$ matrices 
$\Gamma_A=\tau_1^{A_1}\tau_2^{A_2}\tau_3^{A_3}$ and
$B_A=(-\tau_1)^{A_1}(-\tau_2)^{A_2}(-\tau_3)^{A_3},$ 
we carry out a unitary transformation to new fields
\begin{equation}
\begin{array}{cc}
u^{\alpha a p}_i(y)=\frac{1}{4\sqrt{2}}
u^{\alpha a p}_i(y)=\frac{1}{4\sqrt{2}}
\sum_A\Gamma^{\alpha a p}_A\chi_i(A,y), &
d^{\alpha a p}_i(y)=\frac{1}{4\sqrt{2}}
\sum_AB^{\alpha a p}_A\chi_i(A,y),
\end{array}
\end{equation}
where $y$ denotes a site on the ``blocked'' lattice with spacing $2a$ and $A$
labels a site on a $2^3$ cube with its origin at site $y$. 
Each site $x$ of the original lattice 
corresponds, therefore, to a unique choice of $y$ and $A$. The
combination
\begin{equation}
q_i^{\alpha a p}(y)= \left(\begin{array}{c}
u^\alpha_i(y) \\ d^\alpha_i(y)\end{array}\right)^{a p},
\end{equation}
may then be interpreted as a fermion field with spinor index $\alpha$, 
``taste''
index $a$ and isospin index $p$.

We identify the pressure with the negative of the free energy
density $-f=(T/V_s)\ln {\cal Z}$, where $V_s$ is the spatial volume of
the lattice. 
In the large-$N_f$ limit fluctuations in the bosonic fields are
ignored and the partition function is given by
\begin{eqnarray}
{\cal Z}&=&\int{\rm d}q{\rm d}\overline q
\exp\left(-\sum_{i=1}^{N}\sum_k \overline{q}_i M^\dagger(k)M(k)q_i\right)
\exp\left(-S_{\rm b o s}\right)
\nonumber\\
&=&\left(\prod_k\det M(k)\right)^{2N}e^{-S_{\rm b o s}},
\end{eqnarray}
where the fermionic action is written in terms of the fields $q$ and 
$\overline q$ recast in Fourier space. The
effective fermion kinetic operator as a function of fermion momentum
$k_\mu$ is
\cite{AnnPhys}
\begin{eqnarray}
\label{Mk}
M(k)&=&\sum_{\mu=1}^3\frac{i}{2}
\left\{
(\gamma_\mu\otimes\mathds{1}_2\otimes\mathds{1}_2)\sin2k_\mu+
(\gamma_4\otimes\tau_\mu^*\otimes\mathds{1}_2)(1-\cos2k_\mu)\right\} 
\nonumber
\\
&&+\left(\mathds{1}_4\otimes\mathds{1}_2\otimes\mathds{1}_2\right)\Sigma,
\end{eqnarray}
and the bosonic action is
\begin{equation}
S_{\rm b o s}=\frac{2L_s^2L_t N\Sigma^2}{g^2}.
\end{equation}
In the tensor products in eq.~(\ref{Mk}) the first matrix acts on
spinor, the second on ``taste'' and the third on
isospinor indices. $\Sigma$ is the mean field equivalent of $\sigma(\tilde x)$.
Because the blocked lattice spacing is $2a$,
the Brillouin zone ranges from $-\pi/2a$ to $\pi/2a$, meaning that
the allowed momenta are
\begin{equation}
\begin{array}{c c c l l}
\displaystyle
k_\mu&=&\frac{\pi n_\mu}{L_s}&n_\mu=0,\pm1,\ldots,\pm L_s/2
& {\rm for}\; \mu=1,2 \\
k_\mu&=&\frac{\pi (2n_\mu+1)}{2L_t}&n_\mu=0,\pm1,\ldots,\pm L_t/2
& {\rm for}\; \mu=3,
\end{array}
\end{equation}
where $a$ has been set to 1.

The determinant of eq.~(\ref{Mk}) is calculated as
\begin{equation}
\det
M(k)=\left[\frac{1}{4}\sum_{\mu=1}^3\left(\sin^22k_\mu+4\sin^4k_\mu\right)
+\Sigma^2\right]^8,
\end{equation}
so the unsubtracted free energy density is therefore
\begin{equation}
f=-\frac{1}{L_s^2L_t}\ln{\cal Z}=\frac{-2N}{L_s^2L_t}\sum_k\ln\det
M(k) +\frac{2N\Sigma^2}{g^2}.
\end{equation}
As was done with the lattice data, we normalise this by subtracting
the zero temperature contribution so that the pressure as a function
of temperature is defined as 
\begin{equation}
P=-(f_T-f_0).
\label{p_T}
\end{equation}
Finally, the Stefan-Boltzmann limit is found by
calculating eq.~(\ref{p_T}) with $\Sigma\to0$, which corresponds to
$T\to\infty$.

\end{document}